\documentclass[%
reprint,
superscriptaddress,
 amsmath,amssymb,
 jap,
]{revtex4-2}
\usepackage[pdftex]{graphicx}
\usepackage{dcolumn}
\usepackage{upgreek}
\usepackage{bm}
\usepackage[colorlinks=true,linkcolor=blue]{hyperref}
\usepackage[utf8]{inputenc}
\usepackage[english]{babel}
\usepackage{xcolor}

\begin{document}

\title{A wide-range semiclassical self-consistent average atom model}

\author{A. S. Polyukhin}
\email{polyukhinas@gmail.com}
\affiliation{Joint Institute for High Temperatures of Russian Academy of Sciences, 13/2 Izhorskaya st., 125412 Moscow, Russia}
\affiliation{Moscow Institute of Physics and Technology, 9 Institutskiy per., 141701 Dolgoprudny, Moscow Region, Russia}
\author{S. A. Dyachkov}
\affiliation{Joint Institute for High Temperatures of Russian Academy of Sciences, 13/2 Izhorskaya st., 125412 Moscow, Russia}
\address{Dukhov Automatics Research Institute, 22 Sushchevskaya st., 127030 Moscow, Russia}
\author{A. A. Malyugin}
\affiliation{Joint Institute for High Temperatures of Russian Academy of Sciences, 13/2 Izhorskaya st., 125412 Moscow, Russia}
\affiliation{Moscow Institute of Physics and Technology, 9 Institutskiy per., 141701 Dolgoprudny, Moscow Region, Russia}
\author{P. R. Levashov}
\affiliation{Joint Institute for High Temperatures of Russian Academy of Sciences, 13/2 Izhorskaya st., 125412 Moscow, Russia}
\affiliation{Moscow Institute of Physics and Technology, 9 Institutskiy per., 141701 Dolgoprudny, Moscow Region, Russia}

\begin{abstract}
Discovery of material properties at extremes, which are essential for high energy density physics development, requires the most advanced experimental facilities, theories, and computations. Nowadays it is possible to model properties of matter in such conditions using the state-of-the-art density functional theory (DFT) or path-integral Monte--Carlo (PIMC) approaches with remarkable precision. However, fundamental and computational limitations of these methods impede their practical usage while wide-range thermodynamic and transport models of plasma are required. As a consequence, an average atom (AA) framework is still relevant today and has been attracting more and more attention lately. The self-consistent field and electron density in an atomic cell is usually obtained using the Thomas--Fermi (TF), Hartree--Fock (HF), Kohn--Sham (KS) approaches, or their extensions. In this study we present the AA model, where semiclassical wave functions are used for bound states, while free electrons are approximated by the TF model with a thermodynamically consistent energy boundary. The model is compared in various regions of temperatures and pressures with the reference data: Saha model for rarefied plasma, DFT for warm dense matter, and experimental shock Hugoniot data. It is demonstrated that a single AA model may provide a reasonable agreement with the established techniques at low computational cost and with stable convergence of the self-consistent field.
\end{abstract}


\maketitle

\section{Introduction}

Wide-range equations of state (EOS) for matter are demanded in various applications of high energy density physics and physics of plasmas~\cite{Fortov:PhysUspekhi:2009}.
Materials undergoing intense laser \cite{Anisimov:PhysUspekhi:1984,Krasyuk:LaserPhys:2016} or particle beam radiation~\cite{Hoffmann:LaserPartBeams:2005,Gnyusov:LaserPartBeams:2016}, a high- or hypervelocity impact~\cite{Altshuler:PhysUsp:1996}, a high current flow in X or Z-pinch~\cite{Shelkovenko:MRE:2018,McCoy:PRB:2017}, or the confinement fusion~\cite{Betti:NatPhys:2016} may propagate a complex path on a phase diagram starting from normal conditions. The results of quantitative simulations of such processes via computational fluid dynamics methods are very sensitive to the models used to evaluate  thermodynamic and transport properties of matter. Such models have been developing for several decades~\cite{Zeldovich:2002}, however their applicability is restricted to certain regions on the phase diagram. Thus the development of more comprehensive approaches is very important~\cite{Bushman:SovPhysUsp:1983,Lomonosov:LPB:2007}.

The construction of a thermodynamically consistent EOS at extremes is complicated by the presence of free electrons due to the thermal or pressure ionization of atomic shells, which leads to the growth of Coulomb interaction. Its effect may be compared with the energy of thermal (classical) motion of electrons or Fermi energy in case of degenerate electrons using a corresponding coupling parameter $\Gamma = \langle E_{\mathrm{pot}} \rangle / \langle E_{\mathrm{kin}} \rangle = \max(\Gamma_c, \Gamma_q)$, where:
\begin{equation}
  \Gamma_c = \frac{e^2 n_e^{1/3}}{k_B T},\quad \Gamma_q = \frac{e^2 n_e^{1/3}}{\varepsilon_F}, 
  \label{eq:gamma_c}
\end{equation}
$k_B$ is the Boltzmann constant, 
$\hbar$ is the reduced Planck constant, 
$n_e$ is the electron density, 
$m_e$ is the electron mass, 
$T$ is the temperature, 
$e$ is the electron charge, 
$\varepsilon_F$ is the Fermi energy,
\begin{equation}
\varepsilon_F = (3\pi^2)^{2/3}\frac{\hbar^2}{2m_e}n_e.
\end{equation}
Thus, the states of classical or degenerate matter may vary from an ideal system ($\Gamma_{c,q} \ll 1$) to a strongly coupled one ($\Gamma_{c,q} \gg 1$) and a wide-range model should be valid in all corresponding regions of the phase diagram.

Quantum statistical models have been developed in the framework of a continuous electron distribution in a self-consistent field (Thomas--Fermi (TF) model~\cite{Thomas:1927,Fermi:1927} and its extensions~\cite{Feynman:PR:1949,Kirzhnits:SovPhysUsp:1975}) or a direct evaluation of discrete spectrum and wave functions (Hartree--Fock--Slater (HFS) model and its extensions~\cite{Slater:PhysRev:1951,Rozsnyai:PRA:1972,Liberman:PRB:1979,Nikiforov_Novikov_Uvarov:2005}). Nowadays, the most advanced methods to evaluate the properties of electrons in matter are based on the Kohn--Sham density functional theory (KS-DFT)~\cite{Hohenberg:PR:1964}, which provides the electronic structure for an arbitrary distribution of ions~\cite{Kresse:PRB:1993}, so that a variety of properties may be derived, including EOS. However, at relatively high temperatures (hundreds of eV) the use of KS-DFT is hampered as it represents the free electrons with discrete spectrum which results in an overwhelming number of states to compute. Orbital free density functional theory (OF-DFT) represents the non-interacting kinetic energy of electrons as an explicit functional of electronic density~\cite{Clerouin:PRA:1992}, which is a rough approximation at low temperatures where the contribution of single-particle wave functions of bound electrons is essential; thus OF-DFT is physically and computationally acceptable to high-temperature plasmas with free electrons. The combination of OF and KS-DFT may lead to thermodynamic inconsistency, that is why preferable wide-range models should allow a consistent evaluation of the discrete and continuous spectra contributions. This requirement is satisfied in path-integral Monte Carlo (PIMC) models~\cite{Militzer:PRL:2000,Filinov:PRL:2001}, but the computations may be too expensive compared to KS or OF-DFT.

The implementation of multi-ion DFT algorithms with a proper evaluation of electron transitions from bound to free states is quite complicated. The single-ion limitation used in the average atom (AA) framework is more preferrable at developing various approaches for electron structure predictions. AA models have been developing for about a hundred years, so that many improvements have been introduced to the original approach of Thomas and Fermi: quantum and exchange effects~\cite{Dirac:1930,Kirzhnits:JETP:1957,Kirzhnits:JETP:1959,Kalitkin:1960}, shell effects~\cite{Zink:PR:1968,Kirzhnits:JETP:1972,Shpatakovskaya:SovPhysUsp:2012}, binding energy corrections~\cite{Scott:1952} and others. The enhanced TF model allows one to obtain more realistic thermodynamic properties of electrons, but the approximate treatment of the electronic discrete spectrum as the continuous one limits the applicability of this model significantly~\cite{Dyachkov:PhysPlasmas:2014,Dyachkov:PhysPlasmas:2016}.

Modern wide-range AA models should account for bound and free electrons in the self-consistent field, and there are several approaches under development nowadays. It is worth noting the Purgatorio model~\cite{Wilson:JQSRT:2006,Sterne:HEDP:2007}, which is the successor of the Inferno model~\cite{Liberman:PRB:1979}, the superconfiguration approximaion~\cite{Pain:JQSRT:2006}, the HFS-based models~\cite{Nikiforov_Novikov_Uvarov:2005,Sinko:HEDP:2013,Ovechkin:HEDP:2016}, KS-based AA models~\cite{CallowHansen:2022}, and the modern Tartarus code~\cite{Gill:HEDP:2017,Starret:CPC:2019} which is based on the quantum Green's function evaluation~\cite{Starrett:HEDP:2015}. In general, the considered AA models are less computationally expensive compared to multi-ion DFT or PIMC, giving the possibility to construct large tables of plasma properties for practical use. This explains the growing interest to AA models presented in recent studies~\cite{Shemyakin:ContPlasmaPhys:2012,Pain:ContPlasmaPhys:2012,Murillo:PRE:2013,Kadatskiy:HEDP:2019,Orlov:MRE:2019,Faussurier:2021,Aguiar:2021}. AA models may also be extended to include the contribution of ions with a pseudo-atom approach~\cite{Starret:PRE:2016,Ovechkin:HEDP:2019}.

In our study we introduce the AA model, in which semiclassical wave functions for a spherically symmetric atomic cell are used instead of precise ones. The proper joining of discrete and continuous spectra~\cite{Nikiforov_Novikov_Uvarov:2005} provides the thermodynamic consistency, which is required for wide-range EOSs. The use of semiclassical approximation instead of the direct solution of the Shr\"odinger or Kohn--Sham equations makes the model computationally efficient, while keeping a correct electronic structure and robust wide-range convergence at the self-consistent field evaluation. The obtained EOS is compared to the chemical model of rarefied plasma, to KS-DFT in the region of warm dense matter (WDM), and to the TF model at high densities. 

\section{Model formulation}

\subsection{Self-consistent field}
An AA approach suggests that electrons are distributed within a spherical cell of a radius $r_0$ with a nucleus at its center. The electrical neutrality of the cell assumes that the number of electrons equals to the nucleus charge $Z$; the total charge density outside the cell is supposed to be zero. Under these limitations, the electrostatic potential $U$ arising inside the cell can be determined from the Poisson equation:
\begin{equation}
 \Delta U = 4\pi [n_e (\mathbf{r}) - n_i (\mathbf{r})],
  \label{ZTTF:eq:poisson}
\end{equation}
where $n_e$ is the electron density, $n_i = Z\delta(\mathbf{r})$ is the ion density, $\delta(\mathbf{r})$ is the delta function. Another suggestion is the spherical symmetry of the potential and the electron density, so that by applying the appropriate boundary conditions one obtains the following boundary value problem:
\begin{equation}
 \left\{
  \begin{aligned}
    &\frac{1}{r} \frac{\mathrm{d}^2}{\mathrm{d}r^2} \left(r U\right) = n_e(r), \\
    &\left.rU(r)\right|_{r = 0} = Z, \quad U(r_0) = 0, \\
    &\left.\frac{dU}{dr}\right|_{r = r_0} = 0.  \\
  \end{aligned}
  \right.
   \label{eq:boundary_problem}
\end{equation}

The electron-electron correlation and exchange interaction effects may be taken into account using the exchange-correlation functional, which is calculated in the local density approximation based on the formula proposed by Kohn and Sham~\cite{KS:1965}: 
\begin{equation}
\label{eq:change_cor_func}
\Omega_{xc}[n] = \int_{0}^{r_0} 4 \pi r^2 F_{xc}(n(r)) \mathrm{d}r,
\end{equation}
so that the corresponding term 
\begin{equation}
U_{xc}(r) = \frac{\delta \Omega_{xc} }{\delta n(r)} = \frac{\partial F_{xc}(n(r))}{\partial n(r)},
\end{equation}
is added to the potential
\begin{equation}
U_\text{eff}(r) = U(r) - U_{xc}(r).
\end{equation}
It is worth noting that more complex finite-temperature exchange-correlation functionals (including those with an explicit temperature dependence) could be used to study thermodynamic properties~\cite{Ichimaru:PhysReports:1987,Perrot:PRB:2000,Karasiev:PRL:2014}. Probably, they could better describe the relatively low-temperature region (where the free electrons contribution is still small) in comparison with zero-temperature approximations.

In this study we intend to provide an example of using exchange-correlation functionals with our model, so that only a simple zero-temperature one is considered. The presented general AA model suggests that the potential $U_\text{eff}(r)$ depends on the electron density $n_e(r)$ and vice versa.
Using additional assumptions about $n_e(r)$ one can obtain an explicit dependence $n_e(U)$: the well-known TF approximation~\cite{Feynman:PR:1949, Kalitkin:1960} allows one to compute a rough solution consistent with the electroneutrality condition. In a general case, one should find a solution to the problem~\eqref{eq:boundary_problem} iteratively, by applying gradual changes to the potential and electron density until the desired smallness of energy or chemical potential residual is achieved~\cite{Nikiforov_Novikov_Uvarov:2005}; the TF approximation can be used as a first step of the iterative procedure.

\subsection{Discrete and continuous spectra}
The electron density of an AA can be represented as a sum of discrete $n_{d}(r)$ and continuous $n_{c}(r)$ parts. The discrete part is calculated as a quantum-statistical sum over single-electron states, while the continuous part can be obtained using the TF approximation~\cite{Nikiforov_Novikov_Uvarov:2005}:
\begin{equation}
  n_e(r) = n_{d}(r)+n_{c}(r), 
  \label{eq:electron_density}
\end{equation}
\begin{equation}
  n_{d}(r) =\frac{1}{4\pi r^2}\sum_{\varepsilon_{nl} < \varepsilon_0} \frac{2(2l + 1)}
	{1 + \exp\left[(\varepsilon_{nl} - \mu)/T)\right]} R^2_{nl}(r), 
  \label{eq:electron_density_discrete}
\end{equation}
\begin{multline}
n_{c}(r) = \frac{2\sqrt{2}}{\pi^2}\int\limits_{\varepsilon_0}^{\infty} \frac{\sqrt{\varepsilon + U(r)} \mathrm{d}\varepsilon}{1 + \exp\left[(\varepsilon - \mu)/T\right]} = {}\\ 
\frac{(2T)^{\frac{3}{2}}}{2 \pi^2}\left[ I_{\frac{1}{2}}\left(\frac{U(r)+\mu}{T}\right) - I^{inc}_{\frac{1}{2}}\left(\frac{U(r)+\mu}{T}, y^*\right) \right].
\end{multline}
Here $n$ and $l$ are the principal and orbital quantum numbers, respectively, $R_{nl}(r)$ is the radial wave function of the $\{n, l\}$ state, $I_{\frac{1}{2}}$ and $I^{inc}_{\frac{1}{2}}$ are the complete and incomplete Fermi--Dirac integrals~\cite{Cody:MathComp:1967}, respectively, $y^{*} = \max\left\{(\varepsilon_{0} + U(r))/T,0 \right\}$, $\mu$ is the chemical potential, $\varepsilon_{0}$ is the boundary energy between the discrete and continuous states. It is also convenient to express the occupation numbers explicitly:
\begin{equation}
	N_{nl} = \frac{2(2l + 1)}
	{1 + \exp\left[(\varepsilon_{nl} - \mu)/T)\right]}.
\end{equation}

The total number of electrons is determined by the sum of the numbers of occupied states at given energies $\varepsilon_{nl}$ for some chemical potential:
\begin{multline}
	\label{eq:number_of_states_discrete}
	N(\mu) = \int\limits_0^{r_0}4\pi r^2 n(r) \mathrm{d}^3r = \int\limits_0^{r_0}4\pi r^2 [n_d(r) + n_c(r)] \mathrm{d}^3r\\ {} = \sum_{\varepsilon_{nl} < \varepsilon_0}N_{nl}(\mu) + \int\limits_{\varepsilon_0}^{\infty}N_{\varepsilon}(\mu) \mathrm{d}\varepsilon.
\end{multline}
As the spherical cell is electrically neutral, $N(\mu)$ should satisfy the equation $N(\mu) = Z$ which allows to determine an appropriate chemical potential $\mu$.

The boundary between the discrete and continuous spectra $\varepsilon_0$ should be chosen from the condition of the minimum of the grand thermodynamic potential~\cite{Sinko:TVT:1983,Nikiforov:TVT:1987,Nikiforov_Novikov_Uvarov:2005,Pain:JPhysB:2007}:
\begin{multline}
\frac{8\sqrt{2}}{3\pi}\int_0^{r_0}\left[\max\{0,\varepsilon_0+U(r)\}\right]^{\frac{3}{2}}r^2\mathrm{d}r = \\ \sum_{\varepsilon_{nl} <\varepsilon_0} 2(2 l + 1).
   \label{eq:boundary_energy}
\end{multline}
This boundary depends on the thermodynamic conditions of matter (density and temperature) which guarantees the thermodynamic consistency of the model~\cite{Sinko:DocDiss:2005,Pain:JPhysB:2007}. 

\subsection{Wave functions}

Wave functions and energy levels in the AA framework are usually calculated via the one-electron Shr\"odinger equation~\cite{Joshua_Izaac:2018}, which is the most time-consuming part in a simulation. The current study applies a semiclassical approach by Jeffreys, Wentzel, Kramers, and Brillouin (JWKB method), which allows to avoid complicated algorithms with providing very realistic wave functions. The justification of such an approximation can be found in the book by Nikiforov et al.~\cite{Nikiforov_Novikov_Uvarov:2005}, where the difference between exact and approximate wave functions is almost indistinguishable. In this approximation, the radial part of a wave function can be expressed through the Bessel functions:
\begin{multline}
	R^{(i)}_{nl}(r) = \\ 
  \left\{
	\begin{aligned}
	& \frac{C_i}{\sqrt{3}}\sqrt{\frac{\xi_i}{|p|}}K_\frac{1}{3}(\xi_i) & (r \leq r_i), \\
	& \frac{C_i}{\pi}\sqrt{\frac{\xi_i}{p}}\left[J_{-\frac{1}{3}}(\xi_i) + J_\frac{1}{3}(\xi_i)\right] & (r_i \leq r < r_o),
	\end{aligned}
	\right.
	\label{eq:wave_function_inner}	
\end{multline}
\begin{multline}
	R^{(o)}_{nl}(r) = \\ 
  \left\{
	\begin{aligned}
	& \frac{C_o}{\pi}\sqrt{\frac{\xi_o}{p}}\left[J_{-\frac{1}{3}}(\xi_o) + J_\frac{1}{3}(\xi_o)\right] & (r_i < r \leq r_o), \\
	& \frac{C_o}{\sqrt{3}}\sqrt{\frac{\xi_o}{|p|}}K_\frac{1}{3}(\xi_o) & (r \geq r_o).
	\end{aligned}
	\right.
	\label{eq:wave_function_outer}
\end{multline}
Here $J_{-1/3}(x)$, $J_{1/3}(x)$ are the Bessel functions of the first kind, $K_{1/3}(x)$ is the modified Bessel function of the second kind, $o$ and $i$ denote the outer and inner turning points, respectively,
\begin{equation}
	\xi_i(r) = \left|\int\limits_{r_i}^{r}|p_{nl}(r')|\mathrm{d}r'\right|,\  
	\xi_o(r) = \left|\int\limits_{r_o}^{r}|p_{nl}(r')|\mathrm{d}r'\right|.
\end{equation}
The semiclassical radial wave functions oscillate in the region of classical motion $(r_i < r < r_o)$, and exponentially decay in the forbidden regions $r < r_i$ and $r > r_o$. The semiclassical momentum is expressed as
\begin{equation}
\label{eq:semiclassical_momentum}
	p_{nl}(r) = \sqrt{2\left[\varepsilon_{nl} + U(r) - \frac{(l + 1/2)^2}{2 r^2}\right]}.
\end{equation}

\begin{figure*}[t]
  \includegraphics[width=0.49\linewidth]{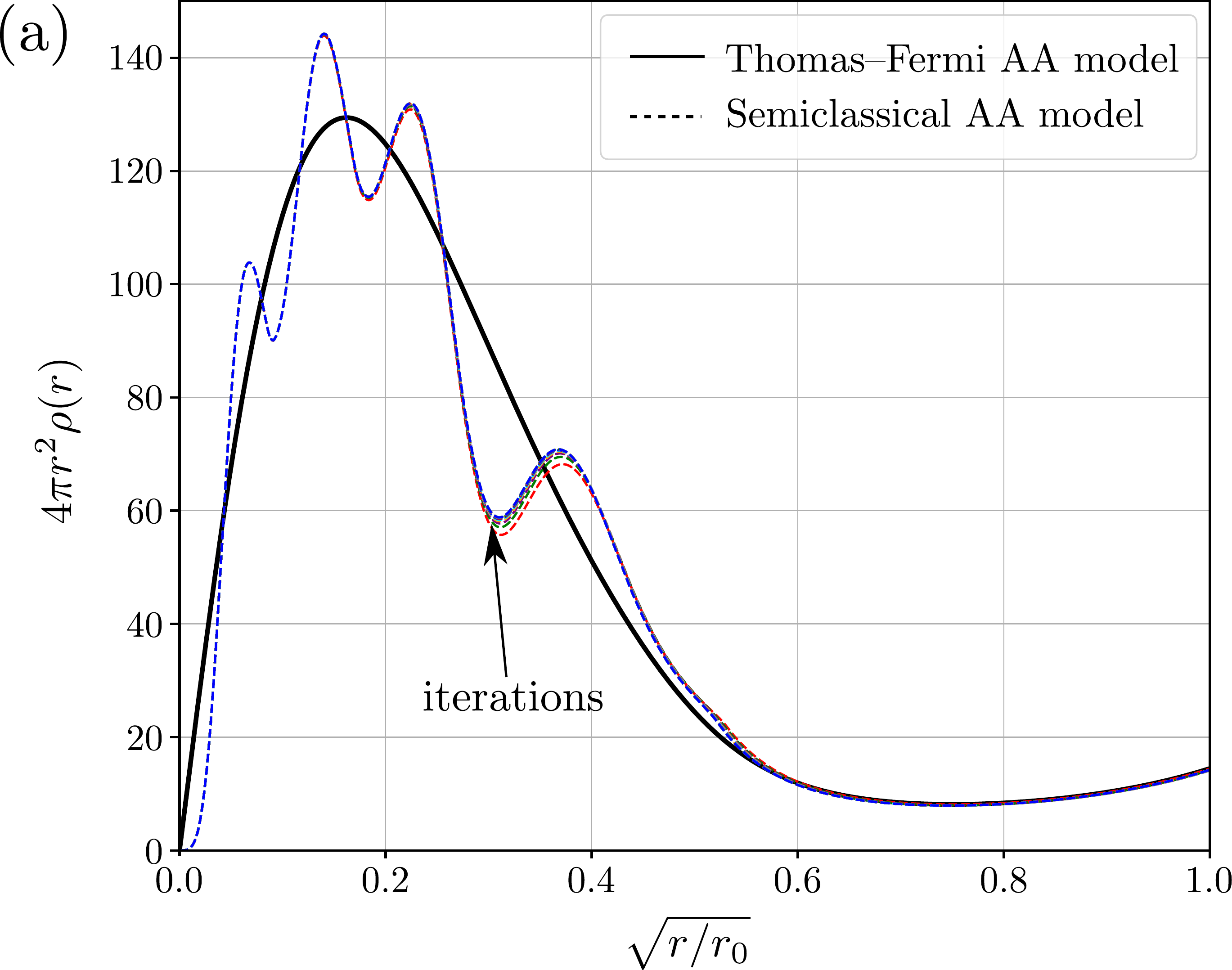}
  \includegraphics[width=0.49\linewidth]{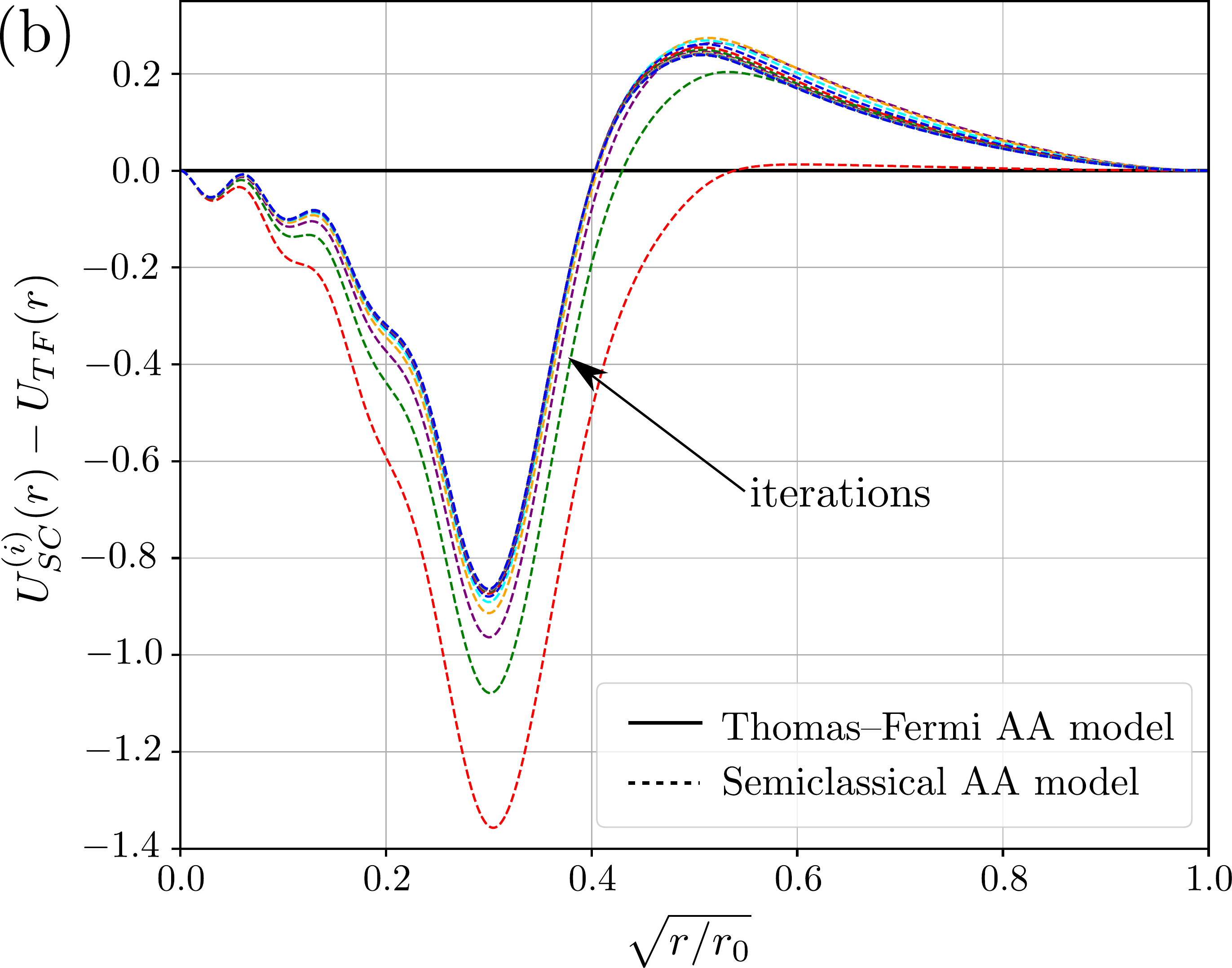}
  \caption{Modeling of the self-consistent field in the atom of gold ($Z = 79$) at $T = 100\,$eV and $\rho = 19.3\,$g/cm$^3$. (a) Evaluation of the radial electron density distribution using the TF model and the developed semiclassical model. The latter produce the electron shells due to accounting for the discrete spectrum. (b) The deviation of the semiclassical potential from the Thomas--Fermi one. The series of dashed lines illustrate the process of convergence to the self-consistent field.}
  \label{fig:image_iterations}
\end{figure*}

The functions $R_{nl}^{(i)}(r)$ and $R_{nl}^{(o)}(r)$ have singularities at the turning points $r_o$ and $r_i$, respectively, which can be removed using the asymptotic expansion near $r_i$ and $r_o$. The smooth joining of the functions \eqref{eq:wave_function_inner},\eqref{eq:wave_function_outer} in the interval $(r_i, r_o)$ is performed as~\cite{Nikiforov_Novikov_Uvarov:2005}:
\begin{eqnarray}
\label{eq:wave-function-complete}
	& R_{nl}(r) = [1 - a(r)]R^{(i)}_{nl}(r) + a(r)R^{(o)}_{nl}(r), \\
	& a(r) = \xi_i(r)/\xi_i(r_o).
\end{eqnarray}
It is worth noting that the signs of the outer wave function $R^{(o)}_{nl}(r)$ and the inner one $R^{(i)}_{nl}(r)$ must be consistent:
\begin{equation}
	C_i = C_o\,\mathrm{sign}\{J_{-1/3}[\xi_i(r_o)] + J_{1/3}[\xi_i(r_o)]\}.
\end{equation}
The value of the constant $C = |C_i| = |C_o|$ is determined by the wave function normalization:
\begin{equation}
	\int_0^{r_0}|R_{nl}(r)|^2\mathrm{d}r = 1.
\end{equation}

\subsection{Energy levels}
The semiclassical energy levels $\varepsilon_{nl}$ are calculated using the Bohr--Sommerfeld quantization condition: 
\begin{equation}
	S_{nl} = \int_{r_i}^{r_o}p_{nl}(r)\mathrm{d}r = \pi (n - l - 1/2).
\label{eq:bohr_sommerfeld_quantization}
\end{equation}
Here $S_{nl}$ is the action integral, $r_i$, $r_o$ are the inner and outer turning points, respectively, $p_{nl}(r)$ is the semiclassical momentum~\eqref{eq:semiclassical_momentum}.

It is well-known that the semiclassical approach is valid only at large enough quantum numbers. For example, Table~\ref{tab:energy} illustrates the difference between the levels calculated using the quantization condition~\eqref{eq:bohr_sommerfeld_quantization} and by solving the Schrödinger equation for the Hartree model. The obtained error may be unacceptable for the spectral properties evaluation, where the accuracy of the energy levels plays a crucial role. But, as it is shown in Section~\ref{sec:results}, the precision of this approximation is enough to describe various thermodynamical effects in a wide range of parameters. 

\begin{table}
	\caption{Energy levels ($\varepsilon_{nl}$ in eV) of gold at $\rho = 0.1\,$g/cm$^3$ and $T = 100\,$eV. The values obtained using the self-consistent Hartree model~\cite{Nikiforov_Novikov_Uvarov:2005} with our semiclassical (SC) approach are shown. Here $n$ and $l$ are the principal and orbital quantum numbers, respectively, $\delta$ is the relative error. }
	\label{tab:energy}
	\begin{center}
			\tabcolsep10pt
			\begin{tabular}{ ccccc } 
				\hline
				$n$ & $l$ & $\varepsilon^{\mathrm{Hartree}}_{nl}$ & $\varepsilon^{\mathrm{SC}}_{nl}$  & $|\delta| ,\%$\\
				\hline\hline
				1 & 0 & -73022  & -72772 & 0.34  \\
				2 & 0 & -12639  & -12517 & 0.97 \\
				2 & 1 & -12151  & -12027 & 1.02 \\
				3 & 0 & -3435.8 & -3369.3 & 1.94 \\
				3 & 1 & -3211.1 & -3141.9 & 2.16  \\
				3 & 2 & -2794.9 & -2715.1 & 2.86  \\
				4 & 0 & -1175.4 & -1184.2 & 0.75 \\
				4 & 1 & -1079.3 & -1088.1 & 0.82  \\
				4 & 2 & -904.68 & -912.94 & 0.91  \\
				4 & 3 & -675.74 & -682.71 & 1.03 \\
				5 & 0 & -543.42 & -552.66 & 1.70  \\
				5 & 1 & -503.02 & -511.97 & 1.78  \\
				5 & 2 & -431.57 & -439.80 & 1.91 \\
				5 & 3 & -344.04 & -351.37 & 2.13 \\
				5 & 4 & -268.24 & -280.98 & 4.75 \\
				\hline\hline
			\end{tabular}
      \tabcolsep10pt
	\end{center}
\end{table}

Figure~\ref{fig:image_iterations} demonstrates the difference between the self-consistent electron density (a) and potential obtained during the convergent iterative procedure (b), and the initial TF approximation for the atom of gold at normal density. The semiclassical electron density has notable oscillations corresponding to the contribution of different electron shells. The complicated electronic structure of gold also affects the self-consistent potential which deviates from the TF potential $U_{TF}(r)$.

\subsection{Thermodynamic functions}

Thermodynamic properties can be obtained by free energy differentiation. Such an approach provides thermodynamic consistency which is an important property for applications. The expression for the free energy $F$ of the AA electron subsystem can be written as a functional of electron density:
\begin{equation}
F[n] = K[n] + U[n] - TS[n] + \Omega_{xc}[n].
\label{eq:free_energy_aa_n}
\end{equation}
Here, $K[n]$ is kinetic energy, $U[n]$ is potential energy, $S[n]$ is entropy, $\Omega_{xc}[n]$ is an exchange-correlation functional. It can be shown that the free energy of the AA model can be rewritten in the following way, convenient for further derivations:
\begin{widetext}
\begin{multline}
\label{eq:free_energy_aa}
F =  - T \sum_{n,l} 2(2l + 1)\ln{ \left[1 + \exp \left( \frac{\mu - \varepsilon_{nl}}{T}\right)\right]} + \frac{1}{2} \int_{0}^{r_0} 4 \pi r^2 n_e(r) \left(U(r) - \frac{Z}{r}\right) \mathrm{d}r + \mu N -\\- \int_{0}^{r_0} 4 \pi r^2 n_e(r) U_{xc}(r) \mathrm{d}r  + \int_{0}^{r_0} 4 \pi r^2 F_{xc}[n_e(r)]\mathrm{d}r.
\end{multline}
By taking the derivatives, the consistent expressions for thermodynamic functions can be obtained:
\begin{multline}
\label{eq:pressure_qc}
P = - \left(\frac{\partial F}{\partial V}\right)_{T} =  -\frac{1}{4 \pi r_0^3} \sum_{\varepsilon < \varepsilon_0} \frac{2l+1}{1+\exp[(\varepsilon_{nl} - \mu)/T]} \Biggr( R_{nl} R^{'}_{nl} + r R_{nl} R^{''}_{nl} -  r  (R^{'}_{nl})^2  \Biggr) \Biggr |_{r= r_0}  +\\ \frac{(2T)^{5/2}}{6 \pi^2} \left[I_{\frac{3}{2}}\left( \dfrac{\mu}{T}\right) - I^{inc}_{\frac{3}{2}}\left( \dfrac{\mu}{T}, y^*\right) \right],  
\end{multline}

\begin{figure*}[t]
\centering
\includegraphics[width=0.49\linewidth]{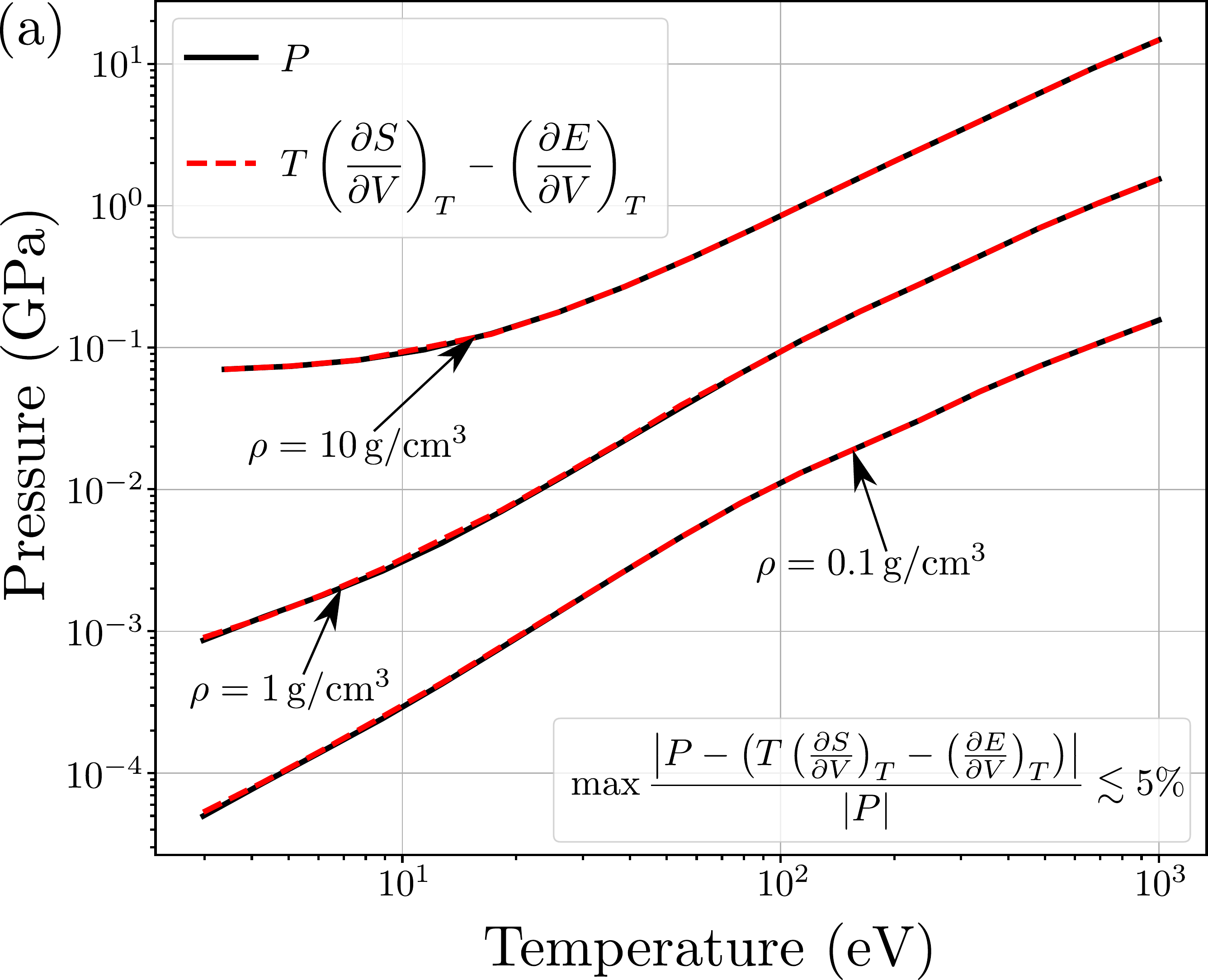}
~
\includegraphics[width=0.49\linewidth]{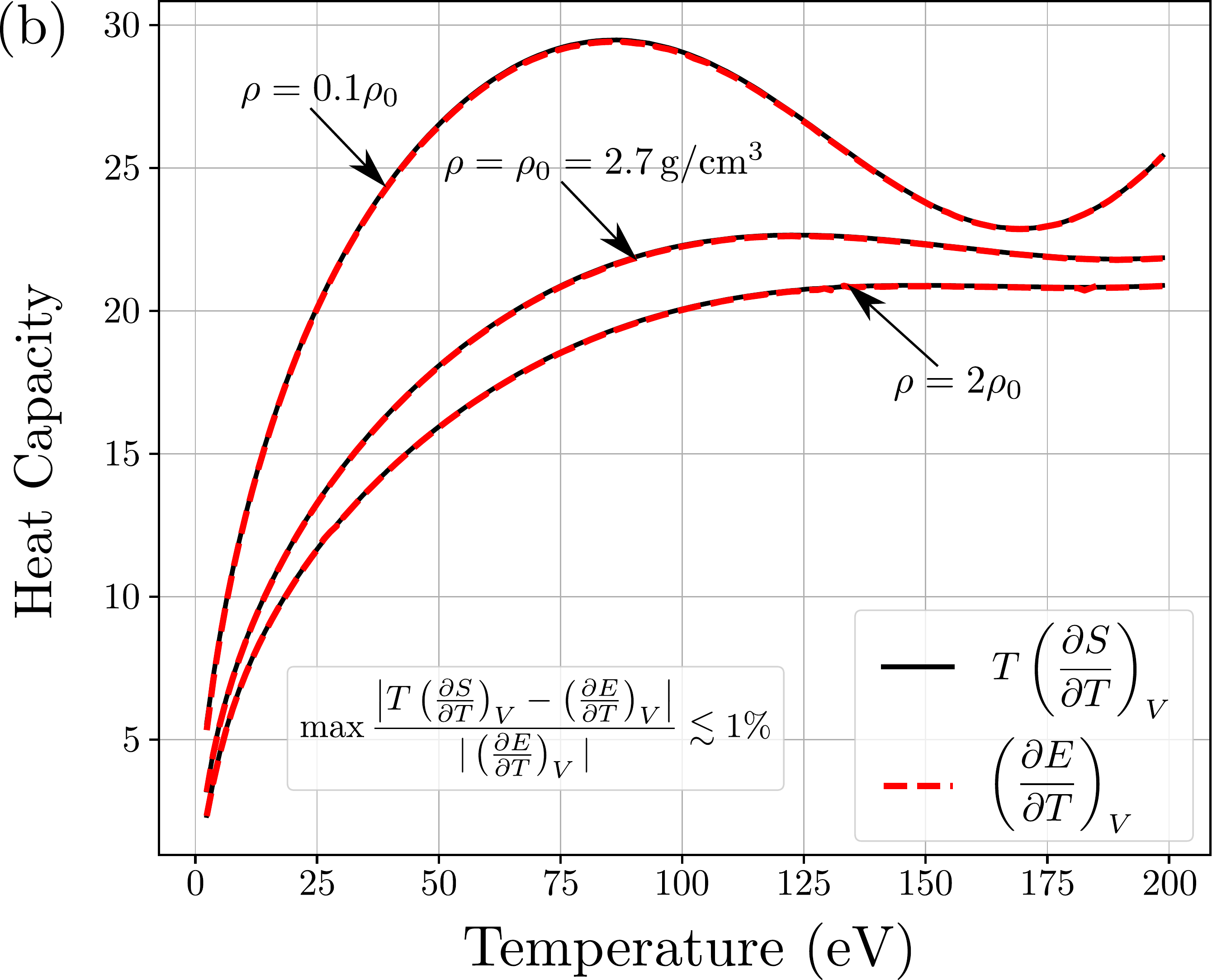}
\caption{Verification of the thermodynamic consistency of the developed semiclassical AA model. (a) Verification of identity~\eqref{eq:consT}: the relative deviation does not exceed $5$\% along the curves. (b) Verification of identity~\eqref{eq:consV}: the relative deviation does not exceed $1$\% along the curves.}
\label{fig:consistency}
\end{figure*}

\begin{multline}
\label{eq:entropy_qc2}
S =  -\left(\frac{\partial F}{\partial T}\right)_{V,N} = -\frac{\mu N}{T} + 2 \sum_{\varepsilon < \varepsilon_0} (2l+1) \ln{\left[1 + \exp\left(\frac{ \mu - \varepsilon_{nl}}{T}\right) \right]}  + \frac{2}{T}\sum_{\varepsilon < \varepsilon_0} \frac{(2l+1)\varepsilon_{nl}}{1+\exp\left[(\varepsilon_{nl} - \mu)/T\right]}  - \\ - \frac{2\sqrt{2}}{3\pi^2 }\ln{\left[1+\exp\left(\frac{ \mu - \varepsilon_0}{T}\right) \right]} \int_{0}^{r_0} 4 \pi r^2 [\varepsilon_0 + U(r)]^{\frac{3}{2}}\mathrm{d}r + \frac{5\sqrt{2}T^{\frac{3}{2}}}{3\pi^2 } \int_{0}^{r_0} 4 \pi r^2 \mathrm{d}r \Biggr[I_{\frac{3}{2}}\left( \dfrac{U(r)+\mu}{T}\right) - \\ I^{inc}_{\frac{3}{2}}\left( \dfrac{U(r)+\mu}{T}, y^*\right) \Biggr] - \frac{1}{T} \int_{0}^{r_0} 4 \pi r^2 U(r) n_c (r) \mathrm{d}r - \int_{0}^{r_0} 4 \pi r^2 \left(\frac{\partial F_{xc}}{\partial T}\right)_{n(r)}\mathrm{d}r,
\end{multline}

\begin{multline}
\label{eq:energy_qc2}
E = F + T S = 2 \sum_{\varepsilon < \varepsilon_0 } \frac{(2l+1)\varepsilon_{nl}}{1+\exp[(\varepsilon_{nl} - \mu)/T]} + \frac{\sqrt{2} T^{5/2}}{ \pi^2}\int_{0}^{r_0} 4 \pi r^2 \mathrm{d}r \left[I_{\frac{3}{2}}\left( \dfrac{U(r)+\mu}{T}\right) - I^{inc}_{\frac{3}{2}}\left( \dfrac{U(r)+\mu}{T}, y^*\right) \right] + \\ +  \int_{0}^{r_0}4 \pi r^2  n_d(r) U (r)  \mathrm{d}r - \frac{1}{2}\int_{0}^{r_0} 4 \pi r^2 \left[U(r) + U_i(r) + U_{xc}(r) \right] n (r) \mathrm{d}r + \\ +\int_{0}^{r_0} 4 \pi r^2 F_{xc}[n(r)] \mathrm{d}r  - T\int_{0}^{r_0} 4 \pi r^2 \left(\frac{\partial F_{xc} [n(r)]}{\partial T}\right)_{n(r)}\mathrm{d}r, 
\end{multline}

\begin{equation}
y^{*} = \max\left(\frac{U(r) + \varepsilon_{0}}{T},0 \right).
\end{equation}
\end{widetext}

\subsection{Thermodynamic consistency}

\begin{figure*}[t]
  \includegraphics[width=0.49\linewidth]{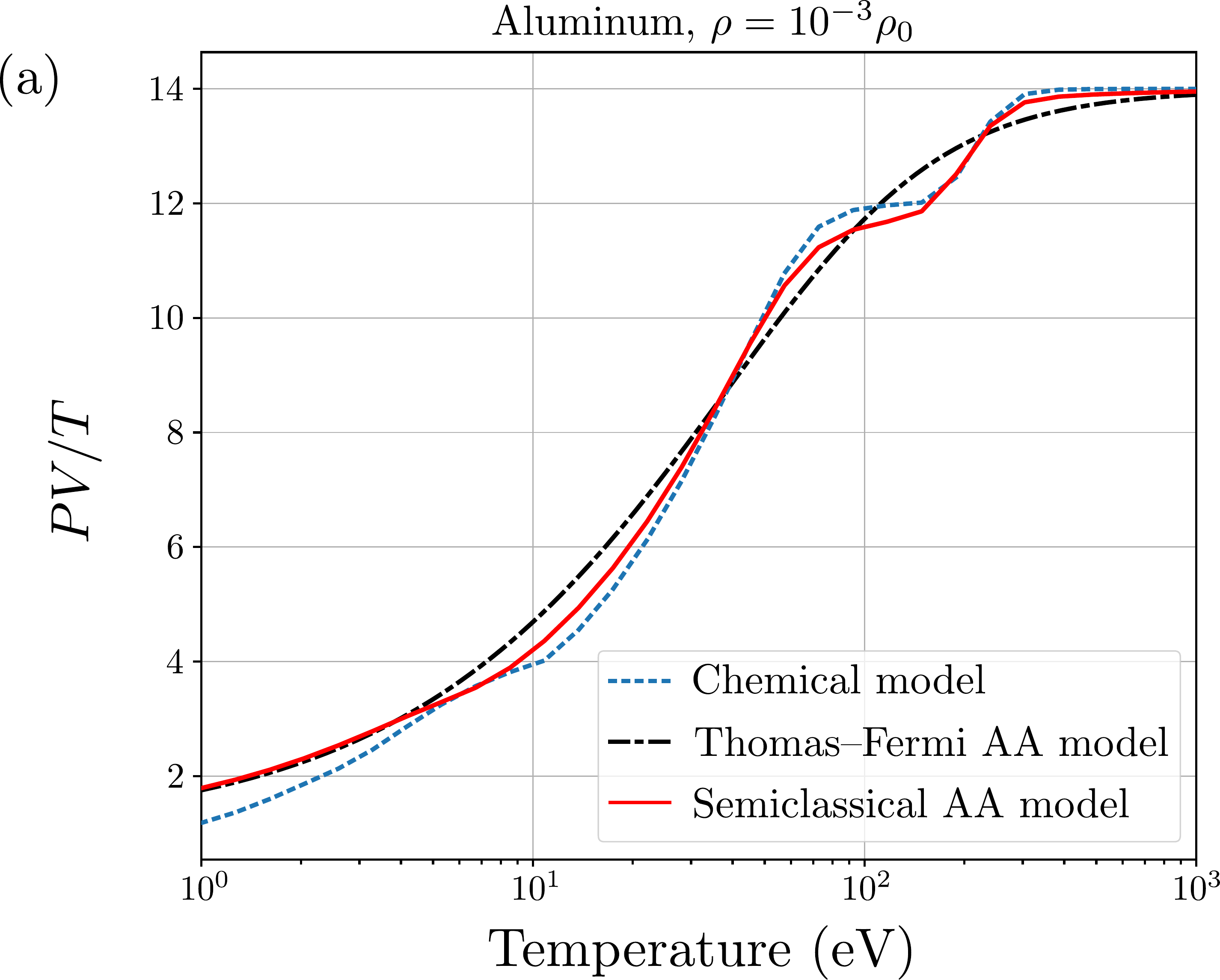}
  \includegraphics[width=0.49\linewidth]{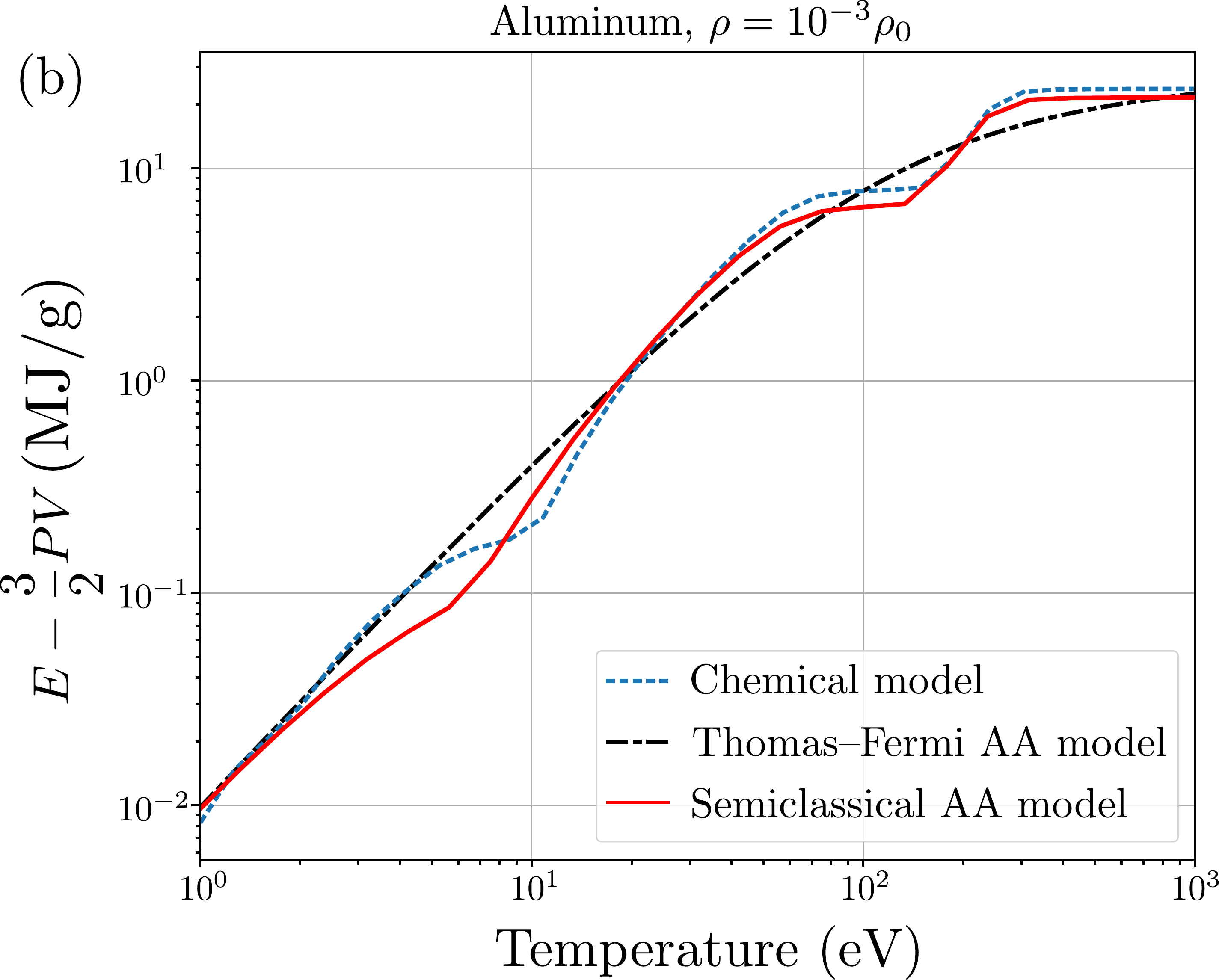}
  \caption{Thermodynamic properties of rarefied aluminum plasma evaluated with the chemical model of plasma (dashed line) and the TF (dash-dotted line) and semiclassical AA (solid line) models for electrons. The ideal gas model for ions was used for the TF and AA models, $\rho = 2.7\times10^{-3}$~g/cm$^3$. (a) The compressibility factor of aluminum. (b) The difference between the energy and kinetic part of energy.}
  \label{fig:eos_saha}
\end{figure*}

The first law of thermodynamics in the differential form $dE = TdS - PdV$ defines two useful constraints between thermodynamic functions. At $T = \text{const}$:
\begin{equation}
  P = T\left(
    \frac{\partial S}{\partial V}
  \right)_T - \left(
    \frac{\partial E}{\partial V}
  \right)_T.
  \label{eq:consT}
\end{equation}
At $V = \text{const}$: 
\begin{equation}
  T\left(
    \frac{\partial S}{\partial T}
  \right)_{V} = \left(
    \frac{\partial E}{\partial T}
  \right)_{V}.
  \label{eq:consV}
\end{equation}
These identities can be used to check the thermodynamic consistency of the AA model.

In Figure~\ref{fig:consistency} we verify Eqs.~\eqref{eq:consT} and \eqref{eq:consV} numerically. The derivatives in \eqref{eq:consT} and \eqref{eq:consV} are calculated with finite differences to demonstrate the convergence of the model. At low temperatures the contribution of the discrete spectrum prevails so the sensibility of results to small changes of temperature or density is higher. Nevertheless, the relative error of identity~\eqref{eq:consT} along isochores does not exceed 5\% in a wide range of temperature, while~\eqref{eq:consV} is satisfied along isotherms even better with the relative error less than 1\%. 

\section{Wide-range evaluation of thermodynamic properties}
\label{sec:results}

The described semiclassical AA model can be applied in a wide range of temperatures and pressures. The validity of the model
can be established by a direct comparison with reference models: at low pressure and density of plasma a chemical model is
precise, for WDM DFT simulations show quite accurate results. The basic TF model is also considered to check
the asymptotic behavior of our semiclassical model at high pressures and temperatures.

\subsection{Low density isochores and chemical model of plasma}

In the low-density region, our semiclassical model is compared to the chemical model of plasma~\cite{Zeldovich:2002,Nikiforov_Novikov_Uvarov:2005} and the finite temperature TF model \cite{Feynman:PR:1949}. The chemical model is based on a set of Saha equations for non-degenerate electron gas:
\begin{multline}
	\label{eq:saha}
	Z_0 \frac{x_{j+1,p}}{g_{j,s}} = {}\\
  \frac{2}{3}\sqrt{\frac{2}{\pi}} r_0^3 T^{\frac{3}{2}} \frac{g_{j+1,p}}{g_{j+1,s}}\exp{\left(-\frac{E_{j+1,p}-E_{j,s}}{T}\right)},
\end{multline}
where $Z_0$ is the average ion charge, $x_{j,s}$ is the fraction of ions with charge $j$ in the state $s$, $g_{j,s}$ is the statistical weight of the corresponding ion state, $E_{j,s}$ is the energy level of the corresponding ion state. Equations~\eqref{eq:saha} are combined with the conditions of electroneutrality:
\begin{equation}
	\label{eq:saha:neutrality}
	\sum_{j,s} x_{j,s} = 1,\
	\sum_{j,s} j x_{j,s} = Z_0.
\end{equation}
Eqs.~\eqref{eq:saha},~\eqref{eq:saha:neutrality} together allow determining concentrations~$x_{j,s}$ for further calculations of thermodynamic properties:
\begin{equation}
	\label{eq:saha:thermodynamic:P}
	P = (Z_0 + 1)\frac{T}{V},
\end{equation}
\begin{equation}
	\label{eq:saha:thermodynamic:E}
	E = \frac{3}{2}(Z_0 + 1)T + \sum_{j,s} x_{j,s}(E_{j,s} - E_{0,0}).
\end{equation}

According to the chemical model, the pressure growth with temperature shown in Fig.~\ref{fig:eos_saha}a has a clear step-wise view caused by the sequential ionization of electron shells. Being thermally ionized, the free electrons contribute to the total pressure as separate particles. However, the shape of the pressure curve is affected much by the electron density distribution within the atomic cell. The characteristic peaks (shells) on the semiclassical density distribution (multiplied by $4\pi r^2$, see Fig.~\ref{fig:image_iterations}) are responsible for the step-wise ionization with the temperature growth, in agreement with the chemical model calculations. The continuous filling of TF electron states results in a smooth distribution of $4\pi r^2\rho(r)$ which does not show characteristic ``pressure steps'' on a low-density isochor during thermal ionization. The latter was noted as one of the major drawbacks of the TF model~\cite{Iosilevsky:TVT:1981}.

Similiar behavior is presented in Fig.~\ref{fig:eos_saha}b for the difference between energy and its kinetic part ($3PV/2$) for aluminum. There is a notable difference between the semiclassical model and the chemical model of plasma near $10\,$eV. On the other hand, at lower and higher temperatures, the models agree better. The reason for such a discrepancy is probably caused by the semiclassical approximation for evaluating the energy spectrum of bound states. 
The energy levels of the outer 3 electrons (3s$^2$ and 3p$^1$), which are ionized first, are calculated with a greater relative error leading to a visible energy deviation compared to the chemical model of plasma.

\subsection{Normal density region with DFT reference}

WDM, which is usually formed under intense laser irradiation or in astrophysical objects, represents partially ionized matter at near-normal density with strong coupling. The most accurate results in this region can be obtained using DFT. Electronic pressure in a dense medium is insensitive to the excitation of inner electrons while the behavior of isochoric heat capacity is largely determined by this effect. Thus, the results for heat capacity will significantly depend on the method of electronic structure calculation.

\begin{figure}[t]
  \center{\includegraphics[width=0.99\linewidth]{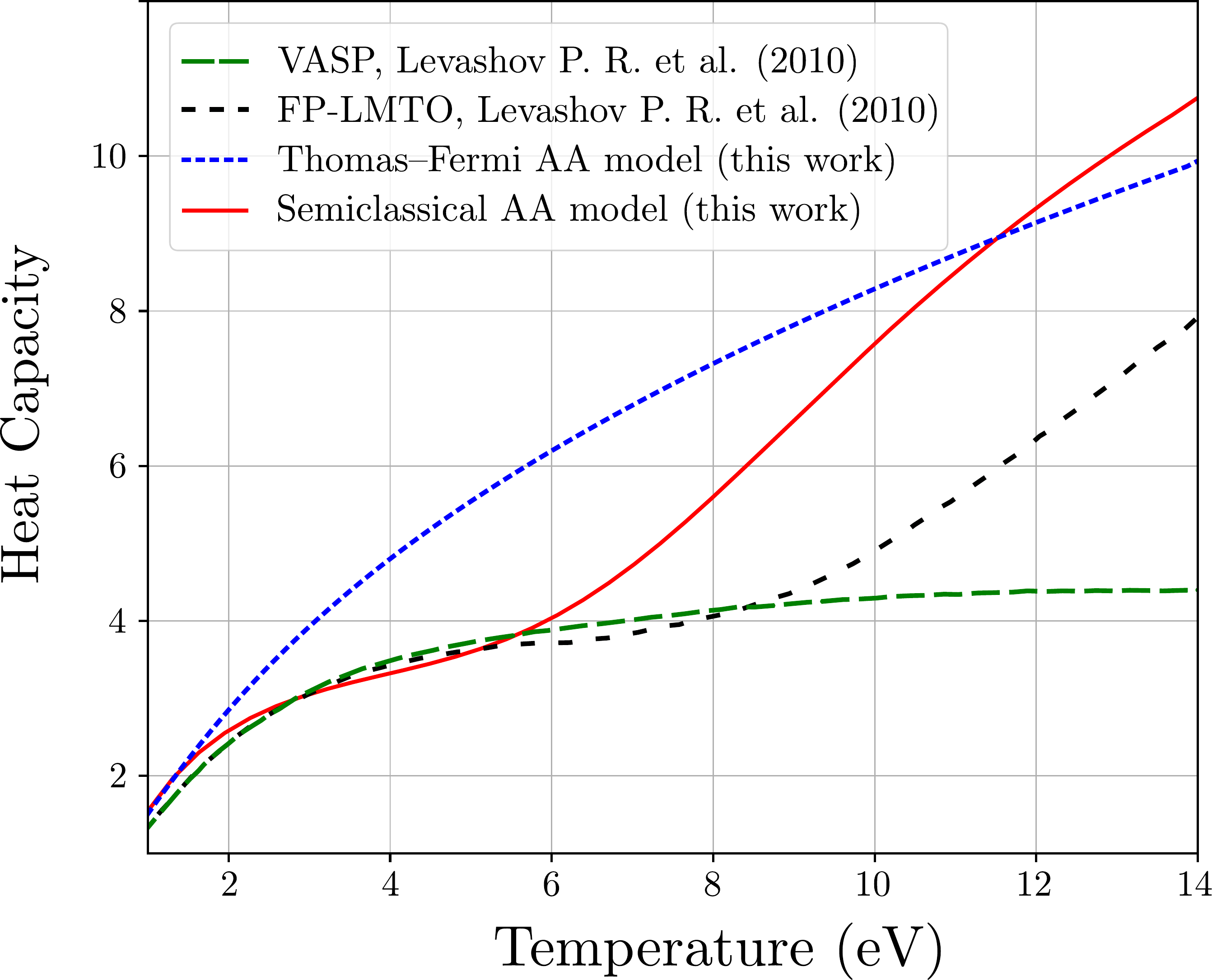}}
  \caption{Isochoric electron heat capacity of aluminum at $\rho = 2.7$~g/cm$^3$. The models are: 1) pseudopotential DFT approach (VASP) with 3s and 3p valent electrons~\cite{Levashov_2010} calculated for crystal; 2) full potential linear-muffin-tin-orbital (FP-LMTO) DFT approach~\cite{Levashov_2010} calculated for crystal; 3) Thomas--Fermi AA model; 4) the developed semiclassical AA model.}
  \label{fig:normal_dens}
\end{figure}

Fig.~\ref{fig:normal_dens} shows the isochoric heat capacity of aluminum calculated via the semiclassical model, the TF model, and the DFT approach~\cite{Levashov_2010}. DFT simulations were performed for crystals using pseudo-potential and full-electron approaches: the first one takes into account only valence electrons, while other electrons form an invariable core with a softer potential for better convergence; the second one considers all electrons and nuclei during a self-consistent field evaluation. As a result, the pseudopotential approach limits heat capacity growth
while other methods behave differently: the semiclassical and full-electron DFT approaches reproduce the sequential ionization of electron shells, and the TF model provides smooth heat capacity growth. There is a difference between the semiclassical model and the full-electron DFT approach in the ionization temperature of lower-level electrons, but both express very similar tendency.

\subsection{High pressure shock Hugoniot evaluation}

Experiments on shock compression are the main source of information for materials at extremes. The Hugoniot equation~\cite{Zeldovich:2002} relates the states of uncompressed and compressed matter and together with an EOS defines the shock Hugoniot curve:
\begin{equation}
    E(V, T) - E_0 = \frac{1}{2}\left(P(V,T)+P_0\right)\left(V_0 - V\right). 
	\label{eq:hugoniot_eq}    
\end{equation}
Here, $P_0$, $V_0$, $E_0$ are the pressure, volume, and specific energy of the uncompressed state (before the shock front).

\begin{figure}[t]
  \center{\includegraphics[width=0.99\linewidth]{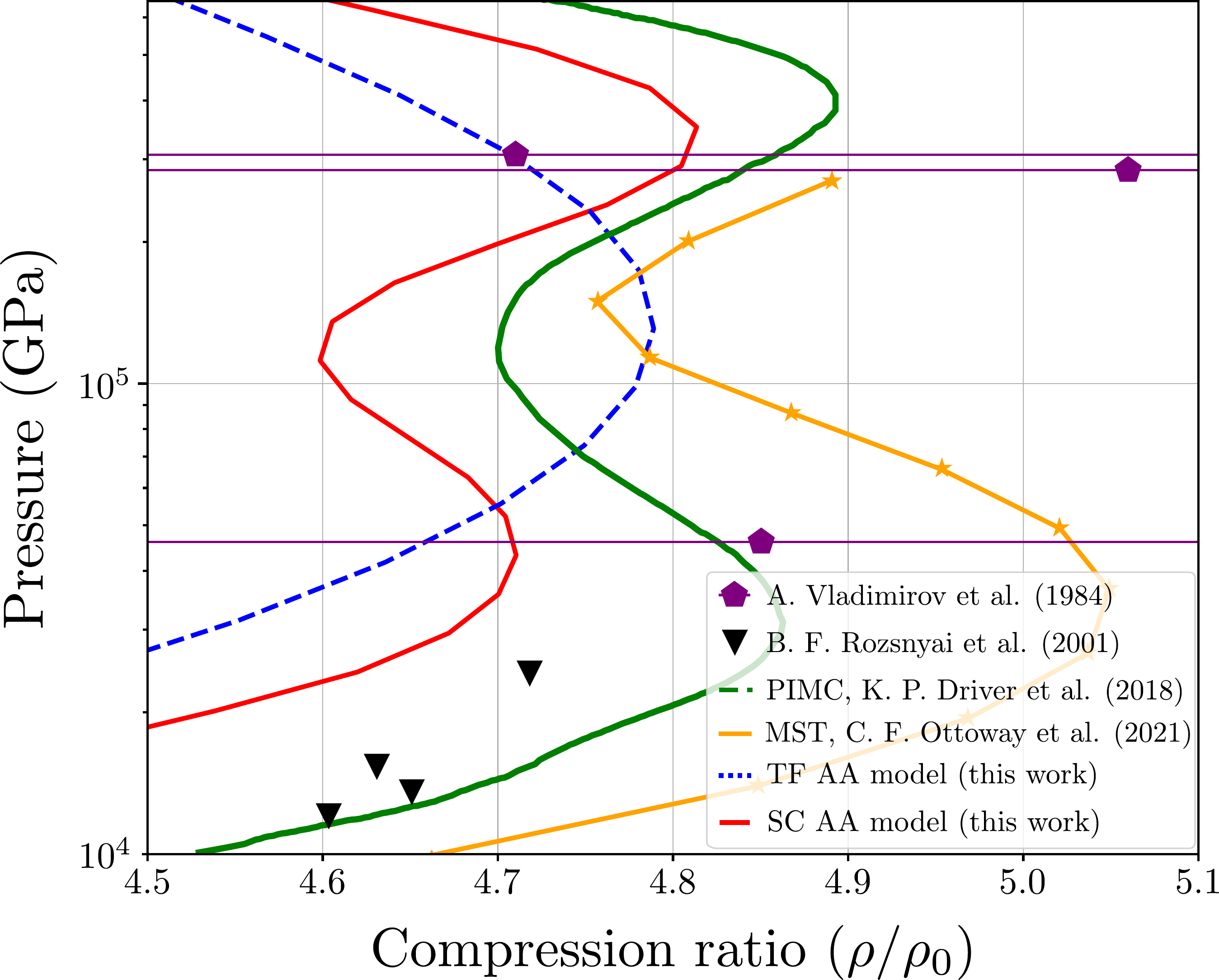}}
  \caption{Shock Hugoniot measurements~\cite{Vladimirov:1984,Rozsnyai:2001} compared to calculations with different models: PIMC model~\cite{Driver:2018} (dashed line), MST model~\cite{Ottoway:PRE:2021} (stars line), Thomas--Fermi AA model (dotted line), and semiclassical AA model (solid line).}
  \label{fig:high_dens}
\end{figure}

Fig.~\ref{fig:high_dens} shows the calculation of shock Hugoniot curve for aluminum using the semiclassical, TF, PIMC~\cite{Driver:2018}, and MST~\cite{Ottoway:PRE:2021} models in comparison with the experimental data~\cite{Rozsnyai:2001,Vladimirov:1984}. An ideal gas model is used for the ions in the semiclassical and TF models. Some experimental points available in the considered region of pressures~\cite{Vladimirov:1984} have huge error bars, so all models fit into the inaccuracy of measurements.

One should note, that the growth of isothermal heat capacity with the thermal ionization of electron shells discussed earlier correlates with the oscillations on the shock Hugoniot. As can be seen, although the ionization effects are well described by the semiclassical AA model, quantitatively it differs from the PIMC and MST results, which provide a greater maximum compression. In particular, this is due to the choice of a simple model describing the contribution of the ions to the EOS. However, even a simple approximation for the ions allows us to show that the semiclassical approach is capable of describing shell effects in agreement with more accurate models. A possible enhancement may include the use of finite-temperature exchange-correlation functionals with our model. Also, the use of a non-relativistic approximation results in the spin degeneracy of electron states, which should be accurately resolved in relativistic models. Such an artificial degeneracy takes effect on the energy levels and, consequently, on the ionization process, which may also shift the maximum level of compression.

\section{Discussion}

 \begin{figure}[t]
  \center{\includegraphics[width=0.99\linewidth]{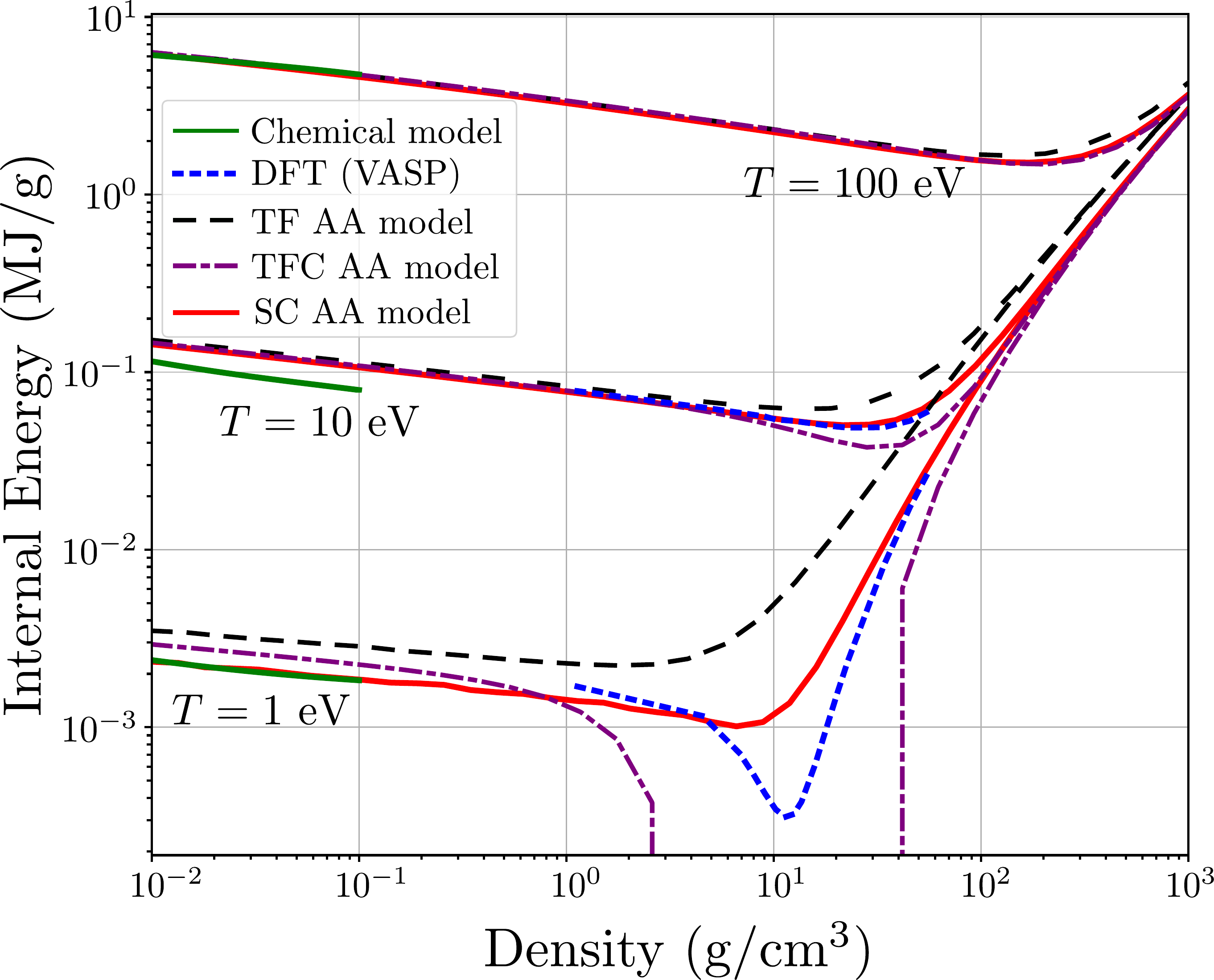}}
  \caption{The wide-range evaluation of the internal energy in lead along isotherms at $1$, $10$, and $100\,$eV. Comparison of the developed semiclassical AA model (SC) with the chemical model of plasma at low densities, DFT calculations (VASP), basic Thomas--Fermi AA model (TF), and the Thomas--Fermi AA model with quantum and exchange corrections (TFC).}
  \label{fig:pb_pressure}
\end{figure}

An example of a wide-range calculation using the semiclassical AA model, along with other models, is demonstrated in Fig.~\ref{fig:pb_pressure}. Pressure on three isotherms for lead (1, 10 and 100~eV) is shown in the range of density from $10^{-2}$~g/cm$^3$ to $10^3$~g/cm$^3$. At low density, the chemical model of plasma agrees quantitatively with the presented approach at 1 and 100~eV. At 10~eV, however, there is a notable overestimation of energy obtained with the semiclassical AA model. As it was mentioned before, this can be connected with a systematic excess of semiclassical energy levels over the corresponding HFS ones. In the region near the  normal density, the semiclassical model predicts the position of the ionization minimum more accurately than the TF one and its modification with quantum and exchange corrections; in the last case the corrections become invalid near the normal density of lead \cite{Dyachkov:PhysPlasmas:2014}. As the temperature rises, the semiclassical model still agrees better with the DFT method than the TF approaches. In the presented diagram TF and TFC models~\cite{Nikiforov_Novikov_Uvarov:2005} can give reasonable results only at high density and high temperatures (in the region of weakly coupled quantum plasma \cite{Dyachkov:PhysPlasmas:2016}, where the continuous electron spectrum determines the values of thermodynamic functions due to ionization. The semiclassical model tends to the TF one in the region of high densities. 

The performance of our calculations using the semiclassical model is worth noting. The self-consistent field evaluation takes the primary calculation time, which is the major contribution to the program's running time. This evaluation itself depends on the number of required energy levels for bound electrons accounting. Depending on the temperature and density, which determine the number of considered levels of the discrete spectrum, the self-consistent field evaluation takes from 10 seconds to several minutes. In addition, parallel programming allows to calculate several points on the phase diagram at once. For example, a modern computer with a 4-core processor can calculate a table of about 250 points with all essential thermodynamic functions of electrons within an hour. Thus, the developed semiclassical AA model can predict the thermodynamic properties of matter with acceptable accuracy at reduced computational costs.

\section{Conclusion}

In this study the semiclassical AA model is introduced. The model considers the discrete and continuous spectrum of electrons in a thermodynamically consistent way.
The use of semiclassical energy levels and wave functions reduces the computational cost while keeping adequate accuracy and provides stable convergence of the self-consistent potential. The combined treatment of bound and free electrons in a single model is essential for modelling ionization and can be used as an integral part of a wide-range AA model.

The results obtained via the semiclassical AA model demonstrate a reasonable agreement with a chemical model of plasma in the region of rarefied plasma, with DFT in the region of WDM, and with shock Hugoniot data. The main drawback of the semiclassical approach is that the energy levels of valence electrons may slightly differ from the exact ones, but it is usually acceptable for a thermodynamic functions evaluation. The presented model may also be used as a part of the pseudo-atom approach~\cite{Starret:PRE:2016} in order to accurately calculate the contribution of ions to plasma properties.

\section*{Acknowledgments}
The authors acknowledge the JIHT RAS Supercomputer Centre, the Joint Supercomputer Centre of the Russian Academy of Sciences,
and the Shared Resource Centre ``Far Eastern Computing Resource''
IACP FEB RAS for providing computing time.


\begin{thebibliography}{10}
\expandafter\ifx\csname url\endcsname\relax
  \def\url#1{\texttt{#1}}\fi
\expandafter\ifx\csname urlprefix\endcsname\relax\def\urlprefix{URL }\fi
\expandafter\ifx\csname href\endcsname\relax
  \def\href#1#2{#2} \def\path#1{#1}\fi

\bibitem{Fortov:PhysUspekhi:2009}
V.~E. Fortov, \href{https://dx.doi.org/10.3367/UFNe.0179.200906h.0653}{Extreme
  states of matter on earth and in space}, Physics Uspekhi 52~(6) (2009)
  615--647.

\bibitem{Anisimov:PhysUspekhi:1984}
S.~I. Anisimov, A.~M. Prokhorov, V.~E. Fortov,
  \href{https://dx.doi.org/10.1070/pu1984v027n03abeh004036}{Application of
  high-power lasers to study matter at ultrahigh pressures}, Physics Uspekhi
  27~(3) (1984) 181--205.

\bibitem{Krasyuk:LaserPhys:2016}
I.~K. Krasyuk, P.~P. Pashinin, A.~Y. Semenov, K.~V. Khishchenko, V.~E. Fortov,
  \href{https://dx.doi.org/10.1088/1054-660x/26/9/094001}{Study of extreme
  states of matter at high energy densities and high strain rates with powerful
  lasers}, Laser Phys. 26 (2016) 094001.

\bibitem{Hoffmann:LaserPartBeams:2005}
D.~H.~H. Hoffmann, A.~Blazevic, P.~Ni, O.~Rosmej, M.~Roth, N.~A. Tahir,
  A.~Tauschwitz, S.~Udrea, D.~Varentsov, K.~Weyrich, Y.~Maron,
  \href{https://dx.doi.org/10.1017/S026303460505010X}{Present and future
  perspectives for high energy density physics with intense heavy ion and laser
  beams}, Laser and Particle Beams 23 (2005) 47--53.

\bibitem{Gnyusov:LaserPartBeams:2016}
S.~F. Gnyusov, V.~P. Rotshtein, A.~E. Mayer, V.~V. Rostov, A.~V. Gunin, P.~R.
  Levashov, \href{https://dx.doi.org/10.1007/s10704-016-0088-8}{Simulation and
  experimental investigation of the spall fracture of 304l stainless steel
  irradiated by a nanosecond relativistic high-current electron beam},
  International Journal of Fracture 199 (2016) 59--70.

\bibitem{Altshuler:PhysUsp:1996}
L.~V. Al'tshuler, R.~F. Trunin, K.~K. Krupnikov, N.~V. Panov,
  \href{https://dx.doi.org/10.1070/pu1996v039n05abeh000147}{Explosive
  laboratory devices for shock wave compression studies}, Physics Uspekhi
  39~(5) (1996) 539--544.

\bibitem{Shelkovenko:MRE:2018}
T.~A. Shelkovenko, S.~A. Pikuz, I.~N. Tilikin, M.~D. Mitchell, S.~N. Bland,
  D.~A. Hammer, \href{https://dx.doi.org/10.1016/j.mre.2018.09.001}{Evolution
  of x-pinch loads for pulsed power generators with current from 50 to 5000
  ka}, Matter Radiat. Extremes 3 (2018) 267--277.

\bibitem{McCoy:PRB:2017}
C.~A. McCoy, M.~D. Knudson, S.~Root,
  \href{https://dx.doi.org/10.1103/physrevb.96.174109}{Absolute measurement of
  the hugoniot and sound velocity of liquid copper at multimegabar pressures},
  Phys. Rev. B 96 (2017) 174109.

\bibitem{Betti:NatPhys:2016}
R.~Betti, O.~A. Hurricane,
  \href{https://dx.doi.org/10.1038/nphys3736}{Inertial-conﬁnement fusion with
  lasers}, Nature Physics 12 (2016) 435--448.

\bibitem{Zeldovich:2002}
Y.~B. Zel'dovich, Y.~P. Raizer, Physics of Shock Waves and High-temperature
  Hydrodynamic Phenomena, Dover Books on Physics Series, Dover Publications,
  2002.

\bibitem{Bushman:SovPhysUsp:1983}
A.~V. Bushman, V.~E. Fortov,
  \href{https://dx.doi.org/10.1070/PU1983v026n06ABEH004419}{Model equations of
  state}, Physics-Uspekhi 26~(6) (1983) 465--496.

\bibitem{Lomonosov:LPB:2007}
I.~V. Lomonosov,
  \href{https://dx.doi.org/10.1017/S0263034607000687}{Multi-phase equation of
  state for aluminum}, Laser and Particle Beams 25 (2007) 567--584.

\bibitem{Thomas:1927}
L.~H. Thomas, \href{https://dx.doi.org/10.1017/S0305004100011683}{The
  calculation of atomic fields}, Mathematical Proceedings of the Cambridge
  Philosophical Society 23~(5) (1927) 542--548.

\bibitem{Fermi:1927}
E.~Fermi, Un metodo statistico per la determinazione di alcune priopriet\`a
  dell'atomo, Rend. Acad. Naz. Lincei 6~(6) (1927) 602--607.

\bibitem{Feynman:PR:1949}
R.~P. Feynman, N.~Metropolis, E.~Teller,
  \href{https://dx.doi.org/10.1103/PhysRev.75.1561}{Equations of state of
  elements based on the generalized {Fermi--Thomas} theory}, Phys. Rev. 75~(10)
  (1949) 1561--1573.

\bibitem{Kirzhnits:SovPhysUsp:1975}
D.~A. Kirzhnits, Y.~E. Lozovik, G.~V. Shpatakovskaya,
  \href{https://dx.doi.org/10.3367/UFNr.0117.197509a.0003}{Statistical model of
  matter}, Soviet Physics Uspekhi 18~(9) (1975) 649.

\bibitem{Slater:PhysRev:1951}
J.~C. Slater, \href{https://dx.doi.org/10.1103/PhysRev.81.385}{A simplification
  of the hartree-fock method}, Phys. Rev. 81 (1951) 385--390.

\bibitem{Rozsnyai:PRA:1972}
B.~F. Rozsnyai, \href{https://dx.doi.org/10.1103/PhysRevA.5.1137}{Relativistic
  hartree-fock-slater calculations for arbitrary temperature and matter
  density}, Phys. Rev. A 5 (1972) 1137--1149.

\bibitem{Liberman:PRB:1979}
D.~A. Liberman,
  \href{https://dx.doi.org/10.1103/PhysRevB.20.4981}{Self-consistent field
  model for condensed matter}, Phys. Rev. B 20 (1979) 4981--4989.

\bibitem{Nikiforov_Novikov_Uvarov:2005}
A.~F. Nikiforov, V.~G. Novikov, V.~B. Uvarov,
  \href{https://dx.doi.org/10.1007/b137687}{Quantum-Statistical Models of Hot
  Dense Matter}, Birkh\"{a}user, Basel, 2005.

\bibitem{Hohenberg:PR:1964}
P.~Hohenberg, W.~Kohn,
  \href{https://dx.doi.org/10.1103/PhysRev.136.B864}{Inhomogeneous electron
  gas}, Phys. Rev. 136~(3B) (1964) 864--870.

\bibitem{Kresse:PRB:1993}
G.~Kresse, J.~Hafner, \href{https://dx.doi.org/10.1103/PhysRevB.47.558}{Ab
  initio molecular dynamics for liquid metals}, Phys. Rev. B 47 (1993)
  558--561.

\bibitem{Clerouin:PRA:1992}
J.~Cl\'erouin, E.~L. Pollock, G.~Zerah,
  \href{https://dx.doi.org/10.1103/PhysRevA.46.5130}{Thomas-fermi molecular
  dynamics}, Phys. Rev. A 46 (1992) 5130--5137.

\bibitem{Militzer:PRL:2000}
B.~Militzer, D.~M. Ceperley,
  \href{https://dx.doi.org/10.1103/PhysRevLett.85.1890}{Path integral monte
  carlo calculation of the deuterium hugoniot}, Phys. Rev. Lett. 85 (2000)
  1890.

\bibitem{Filinov:PRL:2001}
V.~S. Filinov, M.~Bonitz, W.~Ebeling, V.~E. Fortov,
  \href{https://dx.doi.org/10.1088/0741-3335/43/6/301}{Thermodynamics of hot
  dense h-plasmas: Path integral monte carlo simulations and analytical
  approximations}, Phys. Control. Fusion 43 (2001) 743.

\bibitem{Dirac:1930}
P.~A.~M. Dirac, \href{https://dx.doi.org/10.1017/S0305004100016108}{Note on
  exchange phenomena in the {Thomas} atom}, Math. Proc. Camb. Phil. Soc. 26~(3)
  (1930) 376--385.

\bibitem{Kirzhnits:JETP:1957}
D.~A. Kirzhnits, Quantum corrections to the thomas--fermi equation, Soviet
  Journal of Experimental and Theoretical Physics 5 (1957) 115--123.

\bibitem{Kirzhnits:JETP:1959}
D.~A. Kirzhnits, On the limits of applicability of the semiclassical equation
  of state, Soviet Journal of Experimental and Theoretical Physics 35~(6)
  (1959) 1545--1557.

\bibitem{Kalitkin:1960}
N.~N. Kalitkin, \href{http://www.jetp.ac.ru/cgi-bin/dn/e_011_05_1106.pdf}{The
  thomas-fermi model of the atom with quantum and exchange corrections}, Sov.
  Phys. JETP 11 (1960) 1106--1110.

\bibitem{Zink:PR:1968}
J.~W. Zink, \href{https://dx.doi.org/10.1103/PhysRev.176.279}{Shell structure
  and the thomas-fermi equation of state}, Phys. Rev. 176 (1968) 279--284.

\bibitem{Kirzhnits:JETP:1972}
D.~A. Kirzhnits, G.~V. Shpatakovskaya,
  \href{http://www.jetp.ac.ru/cgi-bin/dn/e_035_06_1088.pdf}{Atomic structure
  oscillation effects}, Soviet Physics JETP 35~(6) (1972) 1088--1094.

\bibitem{Shpatakovskaya:SovPhysUsp:2012}
G.~V. Shpatakovskaya, \href{http://ufn.ru/ru/articles/2012/5/a/}{Semiclassical
  model of the structure of matter}, Physics-Uspekhi 55~(5) (2012) 429.

\bibitem{Scott:1952}
J.~M.~C. Scott, \href{https://dx.doi.org/10.1080/14786440808520234}{The binding
  energy of the thomas-fermi atom}, The London, Edinburgh, and Dublin
  Philosophical Magazine and Journal of Science 43~(343) (1952) 859--867.

\bibitem{Dyachkov:PhysPlasmas:2014}
S.~Dyachkov, P.~Levashov, \href{https://dx.doi.org/10.1063/1.4875737}{Region of
  validity of the finite-temperature thomas-fermi model with respect to quantum
  and exchange corrections}, Physics of Plasmas 21~(5) (2014) 052702.

\bibitem{Dyachkov:PhysPlasmas:2016}
S.~A. Dyachkov, P.~R. Levashov, D.~V. Minakov,
  \href{https://dx.doi.org/10.1063/1.4967764}{Region of validity of the
  thomas-fermi model with corrections}, Physics of Plasmas 23~(11) (2016)
  112705.

\bibitem{Wilson:JQSRT:2006}
B.~Wilson, V.~Sonnad, P.~Sterne, W.~Isaacs,
  \href{https://dx.doi.org/10.1103/PhysRevLett.115.176403}{Purgatorio --- a new
  implementation of the inferno algorithm}, J. Quant. Spectrosc. Radiat.
  Transfer 99 (2006) 658.

\bibitem{Sterne:HEDP:2007}
P.~Sterne, S.~Hansen, B.~Wilson, W.~Isaacs,
  \href{https://doi.org/10.1016/j.hedp.2007.02.037}{Equation of state,
  occupation probabilities and conductivities in the average atom purgatorio
  code}, High Energy Density Physics 3 (2007) 278--282.

\bibitem{Pain:JQSRT:2006}
J.~Pain, G.~Dejonghe, T.~Blenski,
  \href{http://dx.doi.org/10.1016/j.jqsrt.2005.05.036}{A self-consistent model
  for the study of electronic properties of hot dense plasmas in the
  superconﬁguration approximation}, J. Quant. Spectrosc. Radiat. Transfer 99
  (2006) 451--468.

\bibitem{Sinko:HEDP:2013}
G.~Sin'ko, N.~Smirnov, A.~Ovechkin, P.~Levashov, K.~Khishchenko,
  \href{https://dx.doi.org/10.1016/j.hedp.2013.02.001}{Thermodynamic functions
  of the heated electron subsystem in the field of cold nuclei}, High Energy
  Density Physics 9~(2) (2013) 309 -- 314.

\bibitem{Ovechkin:HEDP:2016}
A.~A. Ovechkin, P.~A. Loboda, A.~L. Falkov,
  \href{http://dx.doi.org/10.1016/j.hedp.2016.08.002}{Transport and dielectric
  properties of dense ionized matter from the average-atom reseos model}, High
  Energy Density Physics 20 (2016) 38--54.

\bibitem{CallowHansen:2022}
T.~J. Callow, S.~B. Hansen, E.~Kraisler, A.~Cangi,
  \href{https://dx.doi.org/10.1103/PhysRevResearch.4.023055}{First-principles
  derivation and properties of density-functional average-atom models}, Phys.
  Rev. Research 4 (2022) 023055.

\bibitem{Gill:HEDP:2017}
N.~Gill, C.~Starrett,
  \href{http://dx.doi.org/10.1016/j.hedp.2017.06.002}{Tartarus: A relativistic
  green’s function quantum average atom code}, High Energy Density Physics 24
  (2017) 33--38.

\bibitem{Starret:CPC:2019}
C.~Starrett, N.~Gill, T.~Sjostrom, C.~Greeff,
  \href{https://doi.org/10.1016/j.cpc.2018.10.002}{Wide ranging equation of
  state with tartarus: A hybrid green’s function/orbital based average atom
  code}, Computer Physics Communications 235 (2019) 50--62.

\bibitem{Starrett:HEDP:2015}
C.~Starrett, \href{http://dx.doi.org/10.1016/j.hedp.2015.05.001}{A green’s
  function quantum average atom model}, High Energy Density Physics 16 (2015)
  18--22.

\bibitem{Shemyakin:ContPlasmaPhys:2012}
O.~Shemyakin, P.~Levashov, K.~Khishchenko,
  \href{http://dx.doi.org/10.1002/ctpp.201100090}{Equation of state of al based
  on the thomas--fermi model}, Contributions to Plasma Physics 52~(1) (2012)
  37--40.

\bibitem{Pain:ContPlasmaPhys:2012}
J.~Pain, G.~Dejonghe, T.~Blenski,
  \href{http://dx.doi.org/10.1002/ctpp.201100069}{Equation of state of dense
  plasma mixtures: Application to the sun center}, Contributions to Plasma
  Physics 52~(1) (2012) 23--27.

\bibitem{Murillo:PRE:2013}
M.~S. Murillo, J.~Weisheit, S.~B. Hansen, M.~W.~C. Dharma-wardana,
  \href{http://dx.doi.org/10.1103/PhysRevE.87.063113}{Partial ionization in
  dense plasmas: Comparisons among average-atom density functional models},
  Phys. Rev. E 87 (2013) 063113.

\bibitem{Kadatskiy:HEDP:2019}
M.~A. Kadatskiy,
  \href{http://dx.doi.org/10.1016/j.hedp.2019.100700}{Quantum-statistical
  calculations of the thermodynamic properties of molybdenum at high energy
  densities}, High Energy Density Physics 33 (2019) 100700.

\bibitem{Orlov:MRE:2019}
N.~Y. Orlov, M.~A. Kadatskiy, O.~B. Denisov, K.~V. Khishchenko,
  \href{http://dx.doi.org/10.1063/1.5096439}{Application of quantum-statistical
  methods to studies of thermodynamic and radiative processes in hot dense
  plasmas}, Matter and Radiation at Extremes 4 (2019) 054403.

\bibitem{Faussurier:2021}
G.~Faussurier, C.~Blancard, M.~Bethkenhagen,
  \href{http://dx.doi.org/10.1103/PhysRevE.104.025209}{Carbon ionization from a
  quantum average-atom model up to gigabar pressures}, Physical Review E 104
  (08 2021).

\bibitem{Aguiar:2021}
J.~Aguiar, H.~Rocco, F.~Lanzini,
  \href{http://dx.doi.org/10.1140/epjd/s10053-021-00277-3}{Self-consistent
  screened hydrogenic model based on the average-atom model: comparisons with
  atomic codes and plasma experiments}, The European Physical Journal D 75
  (2021) 272.

\bibitem{Starret:PRE:2016}
C.~E. Starrett, D.~Saumon,
  \href{https://dx.doi.org/10.1103/PhysRevE.93.063206}{Equation of state of
  dense plasmas with pseudoatom molecular dynamics}, Phys. Rev. E 93 (2016)
  063206.

\bibitem{Ovechkin:HEDP:2019}
A.~A. Ovechkin, P.~A. Loboda, A.~L. Falkov,
  \href{https://dx.doi.org/10.1016/j.hedp.2019.01.003}{Plasma opacity
  calculations using the starrett and saumon average-atom model with ion
  correlations}, High Energy Density Physics 30 (2019) 29--40.

\bibitem{KS:1965}
W.~Kohn, L.~J. Sham,
  \href{https://dx.doi.org/10.1103/PhysRev.140.A113}{Self-consistent equations
  including exchange and cor relation effects}, Phys. Rev. A. 140 (1965)
  1133--1141.

\bibitem{Ichimaru:PhysReports:1987}
S.~Ichimaru, H.~Iyetomi, S.~Tanaka,
  \href{http://dx.doi.org/10.1016/0370-1573(87)90125-6}{Statistical physics of
  dense plasmas: Thermodynamics, transport coefficients and dynamic
  correlations}, Physics Reports 149~(2) (1987) 91--205.

\bibitem{Perrot:PRB:2000}
F.~Perrot, M.~W.~C. Dharma-wardana,
  \href{http://dx.doi.org/10.1103/PhysRevB.62.16536}{Spin-polarized electron
  liquid at arbitrary temperatures: Exchange-correlation energies,
  electron-distribution functions, and the static response functions}, Phys.
  Rev. B 62 (2000) 16536.

\bibitem{Karasiev:PRL:2014}
V.~V. Karasiev, T.~Sjostrom, J.~Dufty, S.~B. Trickey,
  \href{http://dx.doi.org/10.1103/PhysRevLett.112.076403}{Accurate homogeneous
  electron gas exchange-correlation free energy for local spin-density
  calculations}, Phys. Rev. Lett. 112 (2014) 076403.

\bibitem{Cody:MathComp:1967}
W.~J. Cody, H.~C. {Thacher, Jr.},
  \href{https://dx.doi.org/10.2307/2003468}{Corrigendum: ``{Rational}
  {Chebyshev} approximations for {Fermi}-{Dirac} integrals of orders $-1/2$,
  $1/2$, and $3/2$''}, Math. Comp. 21~(99) (1967) 525--525.

\bibitem{Sinko:TVT:1983}
G.~V. Sin'ko, \href{http://mi.mathnet.ru/rus/tvt/v21/i6/p1041}{Use of the
  self-consistent-field method to calculate the thermodynamic functions of
  electrons in simple materials}, High Temperature 21~(6) (1983) 783--793.

\bibitem{Nikiforov:TVT:1987}
A.~F. Nikiforov, V.~G. Novikov, V.~B. Uvarov,
  \href{http://mi.mathnet.ru/rus/tvt/v25/i1/p12}{The use of semiclassical
  approximation in the modified hartree--fock--slater model}, High Temperature
  25 (1988) 10--19.

\bibitem{Pain:JPhysB:2007}
J.~C. Pain, \href{http://dx.doi.org/10.1088/0953-4075/40/8/008}{A model of
  dense-plasma atomic structure for equation-of-state calculations}, J. Phys.
  B: At. Mol. Opt. Phys. 40 (2007) 1553--1573.

\bibitem{Sinko:DocDiss:2005}
G.~V. Sin'ko, \href{https://search.rsl.ru/ru/record/01003307775}{First
  principles calculations of elastic and thermodynamic properties of matter
  under pressure}, Ph.D. thesis, Federal State Unitary Enterprise "Russian
  Federal Nuclear Center --- Zababakhin All-Russia Research Institue of
  Technical Physics" (2005).

\bibitem{Joshua_Izaac:2018}
J.~Izaac, J.~Wang, Computational Quantum Mechanics, Springer, Basel, 2018.

\bibitem{Iosilevsky:TVT:1981}
I.~L. Iosilevsky, V.~K. Gryaznov,
  \href{http://mi.mathnet.ru/rus/tvt/v19/i6/p1121}{About the precision of
  thermodynamical description of a gas plasma in the thomas--fermi and saha
  approximations}, High Temperature 19 (1981) 1121--1126.

\bibitem{Levashov_2010}
P.~R. Levashov, G.~V. Sin'ko, N.~A. Smirnov, D.~V. Minakov, O.~P. Shemyakin,
  K.~V. Khishchenko,
  \href{https://doi.org/10.1088/0953-8984/22/50/505501}{Pseudopotential and
  full-electron dft calculations of thermodynamic properties of electrons in
  metals and semiempirical equations of state}, Journal of Physics: Condensed
  Matter 22~(50) (2010) 505501.

\bibitem{Vladimirov:1984}
A.~Vladimirov, N.~Voloshin, N.~Nogin, A.~Petrovtsev, V.~Simonenko,
  \href{http://jetpletters.ru/ps/1264/article_19122.pdf}{Shock compressibility
  of aluminum at $p > 1$ gbar}, JETP Lett. 39 (1984) 82.

\bibitem{Rozsnyai:2001}
B.~F. Rozsnyai, J.~R. Albritton, D.~A. Young, V.~N. Sonnad, L.~A.,
  \href{https://doi.org/10.1016/s0375-9601(01)00661-2}{Theory and experiment
  for ultrahigh pressure shock hugoniots}, Phys. Lett. A 291 (2001) 226.

\bibitem{Driver:2018}
K.~P. Driver, F.~Soubiran, B.~Militzer,
  \href{https://doi.org/10.1103/PhysRevE.97.063207}{Path integral monte carlo
  simulations of warm dense aluminum}, Phys. Rev. E 97 (2018) 063207.

\bibitem{Ottoway:PRE:2021}
C.~F. Ottoway, D.~A. Rehn, D.~Saumon, C.~E. Starrett,
  \href{https://link.aps.org/doi/10.1103/PhysRevE.104.055208}{Effect of ionic
  disorder on the principal shock hugoniot}, Phys. Rev. E 104 (2021) 055208.

\end{thebibliography}
\end{document}